\documentstyle[12pt,equation]{article}
\font\elevenbf=cmbx10 scaled\magstep 1
\renewcommand{\theequation}{\arabic{section}.\arabic{equation}}
\newcommand{\k}{\mbox{$|K^0 \rangle$}}
\newcommand{\kb}{\mbox{$|\bar{K^0}\rangle$}}
\newcommand{\ks}{\mbox{$|K_S\rangle$}}
\newcommand{\kl}{\mbox{$|K_L\rangle$}}
\newcommand{\klt}{\mbox{$|K_L(t)\rangle$}}
\newcommand{\kst}{\mbox{$|K_S(t)\rangle$}}
\newcommand{\mss}{\mbox{$\langle K_S \ks$}}
\newcommand{\msst}{\mbox{$\langle K_S \kst$}}
\newcommand{\mll}{\mbox{$\langle K_L \kl$}}
\newcommand{\mllt}{\mbox{$\langle K_L \klt$}}
\newcommand{\msl}{\mbox{$\langle K_S \kl $}}
\newcommand{\mslt}{\mbox{$\langle K_S \klt$}}
\newcommand{\mls}{\mbox{$\langle K_L \ks $}}
\newcommand{\mlst}{\mbox{$\langle K_L \kst$}}
\newcommand{\pkk}{\mbox{$P_{K^0 K^0}(t)$}}
\newcommand{\pbkbk}{\mbox{$P_{\bar{K^0}\bar{ K^0}}(t)$}}
\newcommand{\pkbk}{\mbox{$P_{K^0 \bar{K^0}}(t)$}}
\newcommand{\pbkk}{\mbox{$P_{\bar{K^0} K^0}(t)$}}
\newcommand{\pll}{\mbox{$P_{K_L K_L}(t)$}}
\newcommand{\pss}{\mbox{$P_{K_S K_S}(t)$}}
\newcommand{\psln}{\mbox{$P_{K_S K_L}(t)$}}
\newcommand{\pls}{\mbox{$P_{K_L K_S}(t)$}}

\newcommand{\gsh}{\mbox{$\gamma_{{}_{S}}$}}
\newcommand{\glh}{\mbox{$\gamma_{{}_{L}}$}}
\newcommand{\ml}{\mbox{$m_{{}_{L}}$}}
\newcommand{\ms}{\mbox{$m_{{}_{S}}$}}
\newcommand{\dk}{\mbox{$\Delta_K$}}
\newcommand{\dm}{\mbox{$\Delta m$}}

\begin{document}
\begin{center}
{ \large\bf TIME EVOLUTION OF $K^0-\bar{K^0}$ SYSTEM IN SPECTRAL FORMULATION
\\}
\vskip 2cm
{ Marek Nowakowski \\}
INFN--Laboratori Nazionali di Frascati, P.O. Box 13, I-00044 Frascati
(Roma), Italy
\end{center}
\vskip .5cm
\begin{abstract}
We reanalyse the time evolution of the $K^0-\bar{K^0}$ system in the language
of certain spectral function whose Fourier transforms give the time dependent
survival and transition amplitudes. The reanalysis turned out to be necessary
in view of the astonishing theorem by Khalfin on the possibility of vacuum
regeneration of $K_S$ and $K_L$. The main reason for this unexpected behaviour
is the non-orthogonality of $\ks$ and $\kl$.
As a result of this theorem new contributions
to the well known oscillatory terms will enter the time dependent transition
probabilities. These new terms are not associated with small/large time
behaviour of the amplitudes and therefore their magnitude is apriori unknown.
Approximating the spectral functions by an one-pole ansatz Khalfin
estimated the new effect in transition probabilities to be $4 \times 10^{-4}$.
Whereas we agree with Khalfin on the general existence of vacuum regeneration
of $K_S$ and $K_L$ we disagree on the size of the effect. A careful analysis of
the one-pole approximation reveals that the effect is eleven orders of
magnitude smaller than Khalfin's estimate and, in principle, its exact
determination lies outside the scope of the one-pole ansatz. The present paper
gives also insight into the limitation of the validity of one-pole
approximation, not only for small/large time scales, but also for intermediate
times where new effects, albeit small, are possible. It will be shown that
the same validity restrictions apply to the known formulae of
Weisskopf-Wigner approximation as well.
\end{abstract}

\newpage
\section {Introduction}

The present paper reconsiders an old subject of quantum mechanical
time evolution  of the $K^0-\bar{K^0}$ system. Instead of applying
the well known
Weisskopf-Wigner (WW) approach \cite{ww} to the $K^0-\bar{K^0}$ system
\cite{loy}
we examine the time evolution in the spectral
formalism which is often employed for unstable quantum mechanical
systems \cite{suder1}. In this formulation the Fourier transform of a
spectral density
function gives the time dependent transitions and survival amplitudes. The
reasons to pick up once again
the old subject of time development are twofold. Since the
WW approach is an approximation it is rather useful to have yet another,
different formalism which either confirms the WW results (within a certain
accuracy) or is capable of displaying new (howsoever small) effects.
Due to some
peculiarities of the $K^0-\bar{K^0}$ system one might indeed suspect that
the limitations of the applicability of the WW approximations are, in
principle, different as
compared with other quantum mechanical systems (see below). In view of the
planned high precision experiments in this system it is then
not unreasonable to reconsider this subject. Secondly, the more specific
reason for this reanalysis is a result by Khalfin on the possibilty of vaccum
regeneration of $K_S$ and $K_L$ \cite{kha1}, \cite{kha2}, \cite{kha3}.
The latter would induce new terms in the time
development formulae which according to Khalfin are not completely negligible.
In this paper we investigate this possibilty by using a more refined analysis
than Khalfin's.

The $K^0-\bar{K^0}$ complex is one of the most important test grounds of basic
symmetry properties of nature, like CP- and eventually CPT-(non)conservation
\cite{peccei1}, \cite{peccei2}, \cite{daph1}, \cite{daph2}.
It has also been realized that the $K^0-\bar{K^0}$ system can be used as a
sensitive probe of one of
the fundamental aspect of the theory of nature, namely Quantum
Mechanics \cite{daph1}, \cite{daph2}. This and the fact that the
$K^0-\bar{K^0}$
system is till
now the only system to show
experimental evidence of CP-violation makes it clear why this
specific subject has always played an almost outstanding role in particle
physics. Since the discovery of CP-violation in 1964 \cite{cp}
an enormous number of papers has been devoted to this subject, but even today
it is an alive area and both, the experiment and the theory, try to infer more
information towards a better understanding of CP-violation. We still lack an
experimental confirmation of direct CP-violation (predicted by
the Standard Model) in contrast to the
experimentally established fact of
CP-violation through mixing. Indeed, the two different measurements of
$\Re e(\epsilon'_K /\epsilon_K )$ \cite{re1}, \cite{re2},
one of which \cite{re2} is compatible with zero, are
inconclusive in this respect and further measurements are eagerly awaited (the
ratio $\Im m(\epsilon'_K /\epsilon_K )$  is at present also consistent with
zero \cite{peccei3}). From the theoretical side the basic framework
to calculate
$\epsilon_K$ and $\epsilon'_K$ in the $SU(3)_C \otimes SU(2)_L \otimes U(1)_Y$
Standard Model and its various extensions
is well understood \cite{pasch1}. However, the extraction of exact numbers
predicted from the theory is
still hampered by hadronic uncertainties, like the bag factor,
by poorly known Kobayashi-Maskawa element $V_{ub}$ and the CP-violating
phase $\delta$ as well as by the still
not very precisely known top mass value. Recently further progress has
been made
by realizing the importance of gluonic corrections \cite{pasch2}, \cite{buras}
due to the large top mass and that the latter makes it necessary to include
the contributions of electroweak Penguins \cite{buras}, \cite{roma}.

It is a well appreciated fact that many new models beyond
the Standard Model have
to pass the test of $K^0-\bar{K^0}$ physics putting sometimes severe
restrictions on the model parameters \cite{grimus}.
So, for instance, the {\it general} two
Higgs doublet model which without any further restrictions would predict
flavour changing neutral currents (FCNC)
making the transitions $K^0 \leftrightarrow
\bar{K^0}$ possible at tree level is usually supplemented by a discrete $Z_2$
symmetry to avoid FCNC \cite{pasch3}.
Further examples are the Supersymmetric version of the
Standard Model where off-diagonal gluino-squark-quark coupling gives rise to
a new CP-violating source in the strong sector of the theory which then appears
in a gluino mediated box graph \cite{ray}. An even more `exotic' example of a
$K^0 \leftrightarrow \bar{K^0}$ transition is provided by the R-parity broken
SUSY model where, in principle, this transition can happen through an exchange
of a sneutrino \cite{hall}.
Last but not the least, rare Kaon decays can also put severe limits on
new speculative physics \cite{sehg}
and the connection can go as far as to the, by now
excluded, `fifth force' \cite{nieto}.
All this shows the high sensitivity of the
$K^0-\bar{K^0}$ system.
It is then not a surprise that one can use the $K^0-\bar{K^0}$ physics as a
testing ground of even more speculative assumptions, notably CPT-violation and
violation of Quantum Mechanics.
Eventough both these topics were almost sacrosanct,
recent advances in string theory \cite{ellis}
and formal developments in Quantum Mechanics pioneered by Bell \cite{bell}
made it more
plausible that violation of both might actually ocurr in nature. As far as
CPT-violation is concerned we do not expect the latter to happen in the context
of local, causal and Lorentz-invariant Quantum Field Theories (QFT).
Indeed, the
famous CPT-theorem \cite{cpt}
assures us that with the three aforementioned conditions
CPT is conserved on very general grounds. To circumvent this theorem one has to
drop one of the three underlying conditions. Probably the least painful way
would be to drop the requirement of locality. Such QFT's have been discussed
in the literature, but the status of their consistency is in doubt. String
theories offer possibility to attack this problem: displaying in some sense a
non-local interaction, but being consistent on the other hand. Motivated by the
peculiar feature of Hawking radiation of a Black Hole \cite{hawk}
which allows pure states
to evolve into mixed states, in contradiction with quantum-mechanical results,
a density matrix formalism for the $K^0-\bar{K^0}$ system \cite{hag}
based on string
theory has been developed \cite{ellis} which indeed violates CPT. This would be
CPT-violation through violation of Quantum Mechanics (see also \cite{peskin}).
If such a prediction
comes true it could also be considered as a experimental hint towards string
theories (it is interesting to observe the broad span which connects the
physics of a Black Hole with the physics e.g. at Da$\Phi$ne).
Quite independent how a
possible CPT-violation arises the test of the `last discrete space-time
symmetry' which seems unbroken till now is important (for the status of
CPT-violation from experiment see \cite{erice}).

Also independent of any specific theory a precision experiment of Quantum
Theory in the $K^0-\bar{K^0}$ system is desirable. Doubts about validity of
Quantum Theory in general
date back to the birth of Quantum Mechanics highlighted by
arguments like the EPR-paradox \cite{epr} and speculations about
hidden variable theories \cite{bellbook}.
A general set up of a local realistic models versus Quantum Mechanics has been
reanalysed by Bell \cite{bell}
providing us with the tool of the known Bell-inequalities
which arise in the context of a posible hidden variable theory.
Experiments with
spin correlations show that this inequality is violated \cite{aspect}
and hence QM
confirmed, at least in this case. In ref.\cite{ghi}
a version of Bell-inequalities
has been derived whose examinations revealed that these inequalities are not
violated by QM predictions for any choice of the parameters. However, recently
a proposal has been made to test QM versus a local theory at the $\Phi$-factory
with the help of Bell-inequalities by using $K_L-K_S$
regeneration in {\it matter} \cite{eberh}. A suggested
test of quantum mechanical superposition principle \cite{dass}
can also be counted in the
realm of general tests of QM. For yet different possibilities and developments
we refer the
reader to \cite{datta1}, \cite{sriv}.

It is worth stressing that many ongoing and suggested tests, as well as their
refutals, of CP-, T-, \cite{kabir1}, CPT-symmetry and QM
have directly to do with the
time evolution of the system. This brings us back to the quantum mechanical
time development which is indeed, beside the theoretical determination of the
system parameters $\epsilon_K$ and $\epsilon'_K$, the second pillar of the
$K^0-\bar{K^0}$ system and which is much less model dependent than the latter.
Keeping in mind that any possible violation of CPT and QM is forced to be
rather small it is quite important to examine the nature of new effects the
time development might hide beyond the WW approximation (the WW approach is an
approximation, though a rather good one). To understand why deviations from WW
are expected let us recall a well founded theorem which confirms deviations
from the exponential decay law $\exp (-\Gamma t)$ for very small (the region of
`quantum Zeno' effect \cite{zeno}) and very large times \cite{time}.
It is also known that the
exponential decay law can be derived consistently up to terms of order
$\Gamma / M$ \cite{bohm} which in the case of interest is
\begin{equation} \label{1.1}
{\Gamma_{{}_{X}} \over m_{{}_{X}}} \sim 10^{-15}, \, \, \, X=K_S, K_L
\end{equation}
That the situation in the $K^0-\bar{K^0}$ system might be different can be seen
from the following reasoning. First, due to mixing the mass difference
$\ml -\ms$ will enter the transition probabilities like
$|\langle K^0|\bar{K^0}(t)\rangle |^2$ etc.
We then find, in addition to (\ref{1.1}), other dimensionless
quantities like
\begin{eqnarray} \label{1.2}
&{\Gamma_{{}_{S}} \over m_{{}_{L}}-m_{{}_{S}}} \sim {\cal O}(1), \, \, \,
{\ml - \ms \over \ml} \sim 10^{-15} \nonumber \\
&{\Gamma_{{}_{L}} \over m_{{}_{L}}-m_{{}_{S}}} \sim 10^{-3}
\end{eqnarray}
Of course no new effects will be present which go hand in hand with the first
ratio. It is the third dimensionless ratio in (\ref{1.2}) which is intruguing
and which is small enough to be dropped in the first approximation, but on the
other hand not small enough to be neglected completely.

The second reason why the $K^0-\bar{K^0}$ system differs from a `normal'
unstable quantum mechanical system is that the $K_S$ and $K_L$, defined as
usual, are not orthogonal to each other due to the presence of CP-violation in
the mixing. This peculiar property causes sometimes problems like e.g.
EPR-like paradox \cite{datta2}
and gives rise to questions for the anti-particles of $K_S$ and $K_L$.
For a more detailed discussion on this issue we refer the reader to
the papers \cite{suder2}.
Based on this non-orthogonality Khalfin has proved, in
the formalism of spectral functions $\rho_{{}_{S}}$ and $\rho_{{}_{L}}$
(suitable also otherwise for any unstable quantum mechanical system) that
the {\it vacuum} (in contrast to similar phenomena in {\it matter})
regeneration probability of $K_S \leftrightarrow K_L$ is {\it non-zero}
unless there is no CP-violation through mixing in the $K^0-\bar{K^0}$ system
\cite{kha2}, \cite{kha3}, \cite{suder3}.
To
estimate this effect he uses a reasonable one-pole approximation for
$\rho_{{}_{S}}$ and $\rho_{{}_{L}}$ and finds then indeed {\it new} terms in
the transition probability $|\langle K^0|\bar{K^0}(t)\rangle |^2$ etc.
which are
of the order of $\Gamma_{{}_{L}} /(\ml -\ms )$. We agree with Khalfin on a
general existence of such an effect of vacuum regeneration of $K_S$ and $K_L$
once the $K_S$ and $K_L$ are defined in the usual manner. But we disagree on
the numerical estimate of this effect. It will be shown below that a consistent
treatment of the spectral formalism in general and the one-pole approximation
in specific yields a quite different picture as far as the size of this
`new' effect is concerned. Indeed Khalfin does not use all the information
available in the formalism which, in our opinion, leads to the wrong estimate.
In detail the following will be shown below
\begin{itemize}
\item[(i)]
Taking into account all available information on the spectral functions
$\rho_{{}_{S}}$ and $\rho_{{}_{L}}$ we investigate the consistency of the
one-pole approximation and find that it is valid up to terms of order
$\Gamma_{{}_{X}} /m_{{}_{X}}$, $(\ml -\ms) /\ml$. It will be argued that such
corrections do arise {\it not only} for very large and very small time scales.
\item[(ii)]
Through this consistency check we can determine all parameters of the one-pole
approximation needed for the time evolution formulae (again up to accuracy of
$\Gamma_{{}_{X}} /m_{{}_{X}}$, $(\ml -\ms) /\ml$) in terms of known quantities.
\item[(iii)]
This makes it possible to derive time evolution formulae like
$|\langle K^0|\bar{K^0}(t)\rangle |^2$ etc. in the spectral formalism
and with the one-pole ansatz
(in the accuracy mentioned above) without any further
assumptions. The result of a lengthy calculation is that all formulae agree
with the corresponding expressions derived within the WW approach.
\item[(iv)]
In consequence this result shows explicitly that the vacuum regeneration
probability must be of the order
$\Gamma_{{}_{X}} /m_{{}_{X}}$, $(\ml -\ms) /\ml$. This, however, does not mean
that that such an effect is associated with small/large time behaviour of the
amplitudes.
\end{itemize}

This work was inspired by a talk given by Khalfin in the Second Da$\Phi$ne
Meeting. After the main bulk of the work has been finished the author of the
present paper became aware of a paper by Chiu and Sudershan \cite{suder3}
who treat the same
subject. These authors use the solvable Friedrichs-Lee model to show that, in
general, Khalfin's conclusions on the vacuum regeneration are indeed correct.
However, they disagree with Khalfin's numerical estimate.
The present paper arrives
at the same conclusion as \cite{suder3}
with the difference that no specific model is
needed. It will be shown below that
one can arrive at these conclusions by a careful analysis of the
one-pole ansatz.

The paper is organized as follows. In section 2 we collect all essential and
quite general formulae for the time development. In section 3 we present two
of Khalfin's results. Section 3 investigates the one-pole ansatz and its
consistency. In section 4 all the forgoing results will be gathered to
derive the time evolution of the system. In section 5 we present our
conclusions.

\setcounter{equation}{0}
\section{Basic Formulae}

Out of the Weisskopf-Wigner approximation we will essentially need only the
part which has to do with the eigenvectors of the effective, non-hermitian
Hamiltonian which is the result of two approximations made in the Schr\"odinger
equation \cite{nach}, \cite{kabir3}.
This part defines the $K_S$ and $K_L$ states in the usual way
\begin{eqnarray} \label{2.1}
&\ks = p\k + q \kb, \, \, \, \, \kl = p\k -q\kb  \\
\label{2.2}
&\mss = \mll = |p|^2 + |q|^2 =1 \\
\label{2.3}
&\msl = \mls =|p|^2 - |q|^2 \neq 0
\end{eqnarray}
The equality $\msl =\mls$ in eq.(\ref{2.3}) is imposed by CPT-invariance which
we will assume to hold throughout the paper. The presence of
CP-violation in the mixing
is reflected by $|p|^2 - |q|^2 \neq 0$ which enforces the states $K_S$ and
$K_L$ to be non-orthogonal to each other. Another often used parametrization of
the mixing parameters is given by
\begin{equation} \label{2.4}
p={1+ \epsilon_K  \over \sqrt{2(1 + |\epsilon_K|^2)}},\, \, \, \, \,
q={1- \epsilon_K  \over \sqrt{2(1 + |\epsilon_K|^2)}}
\end{equation}
which makes contact with the $\epsilon_K$ parameter mentioned in the
introduction. Since the CP-violation in the $K^0-\bar{K^0}$ system
(or equivalently the non-orthogonality of $K_S$ and $K_L$) will play
an important role we define for the sake of a short notation
\begin{equation}
\label{2.5}
\dk \equiv |p|^2 - |q|^2
\end{equation}
Let us also note here that although eqs.(\ref{2.1})-(\ref{2.3}) come out
naturally in the context of WW-approximation, indepedent of this
approximation, assuming
$K^0 \leftrightarrow \bar{K^0}$ mixing, the presence of CP-violation in the
mixing and implementing therein CPT-contraints there is
not much choice left other than to postulate
eqs.(\ref{2.1})-(\ref{2.3}) for the $K_S$ and $K_L$ states,
up to possible contributions
from continuum states which we neglect (for a different
point of view where in the context of a generalized quantum mechanical vector
space
$K_S$ and $K_L$ are orthogonal see \cite{suder2} and references therein).
Hence eqs.(\ref{2.1})-(\ref{2.3}) have a much broader applicability than the
part of WW approximation which determines the time dependence of transition
and survival amplitudes.

Given a full, hermitian Hamiltonian $H$ according to general principles of
Quantum Mechanics the time evolution for $K^0$ and $\bar{K^0}$ can be
summarized as follows
\begin{eqnarray} \label{2.6}
&P_{K_{\alpha}K_{\beta}}(t)=\langle K_{\alpha}|e^{-iHt}|K_{\beta}\rangle
=\langle K_{\alpha}|
K_{\beta}(t)\rangle \nonumber \\
&|K_{\alpha}(t)\rangle = e^{-iHt}|K_{\alpha}\rangle \nonumber \\
&K_{\alpha}= K^0,\, \bar{K^0},
\end{eqnarray}
Due to the non-orthogonality of $K_S$ and $K_L$ there is a subtle difference
between the treatment of the time evolution of $K^0$, $\bar{K^0}$ and
$K_S$, $K_L$. For the former the $P_{K_{\alpha}K_{\beta}}(t)$ are expansion
coefficients in
\begin{eqnarray} \label{2.7}
|K^0 (t)\rangle &=& \pkk |K^0 \rangle + \pbkk |\bar{K^0}\rangle \nonumber \\
|\bar{K^0}(t) \rangle &=& \pbkbk |\bar{K^0} \rangle + \pkbk |K^0\rangle
\end{eqnarray}
which according to the orthogonality of $K^0$ and $\bar{K^0}$ and
in agreement with the first equation in (\ref{2.6}) are identical to
$\langle K_{\alpha}|K_{\beta}(t)\rangle$ for $K_{\alpha}=K^0, \, \bar{K^0}$.
Since the quantum mechanical
principle $|A(t)\rangle =\exp (-iHt)|A\rangle$ is valid for any state
$|A\rangle$ we can use eqs.(\ref{2.1})-(\ref{2.3}) and eq.(\ref{2.6}) to derive
the following time dependence of $K_S$ and $K_L$
\begin{eqnarray} \label{2.8}
\kst = p\left[\pkk \k + \pbkk \kb \right] +
q\left[\pbkbk \kb + \pkbk \k\right] \nonumber \\
\klt = p\left[\pkk \k + \pbkk \kb \right] -
q\left[\pbkbk \kb + \pkbk \k \right] \nonumber \\
\end{eqnarray}
Note that in this section we are keeping all formulae as general as possible,
in accordance with the general principles of Quantum Mechanics.
In analogy to eq.(\ref{2.7}) and again in full generality
we can also define
expansion coefficients
$\pss$, $\pll$, $\pls$ and $\psln$ through
\begin{eqnarray} \label{2.9}
\kst = \pss \ks + \pls \kl \nonumber \\
\klt = \pll \kl + \psln \ks
\end{eqnarray}
Clearly the time dependent functions $\psln$ and $\pls$, absent in the WW
approximation, would be, unless identical
to zero, responsible for vacuum regeneration
of $K_S \leftrightarrow K_L$.
Using already the following CPT-constraint (being at same time a quite
model-indendent test for CPT conservation \cite{kha3})
\begin{equation} \label{2.10}
\pkk = \pbkbk
\end{equation}
the $\pss$ etc
can be easily obtained from (\ref{2.8}) by using the inverse transformation of
eq.(\ref{2.1}). The result is
\begin{eqnarray} \label{2.11}
&\pss - \pll ={q \over p} \pkbk + {p \over q} \pbkk \nonumber \\
&\pss + \pll = \pkk + \pbkbk = 2 \pkk \nonumber \\
&\pls = -\psln ={1 \over 2}\left\{{q \over p} \pkbk -{p \over q}\pbkk \right\}
\end{eqnarray}
Trivially eqs.(\ref{2.9}) imply a relation between the expansion coefficients
$\pss$ etc and the corresponding matrix elements $\msst$ etc.
\begin{eqnarray} \label{2.12}
\msst = \pss + \pls \dk \nonumber \\
\mslt = \pss \dk + \pls \nonumber \\
\mllt = \pll - \pls \dk \nonumber \\
\mlst = \pll \dk  - \pls
\end{eqnarray}
This explicitly displays the above mentioned difference between the $K^0$,
$\bar{K^0}$ and the $K_S$, $K_L$ cases. The matrix element e.g. $\mlst$ is not
equal to the corresponding coefficient $\pls$. Only if {\it both},
$\pls=-\psln=0$ {\it and} $\dk =0$, are imposed is
this equality guaranteed. Hence this property, $\mlst \neq \pls$,
has nothing to do with the generality of our formulae, but in general
with the fact that $\dk \neq 0$.

Let us now come to the main point of the paper.
The question which will be addressed in the next sections is whether
\begin{equation} \label{2.13}
\pls = -\psln =0 \, \, \, {\rm or} \, \, \, \neq 0
\end{equation}
As discussed in the introduction Khalfin has proved \cite{kha2}, \cite{kha3}
(confirmed
in \cite{suder3})
that indeed the second possibilty must be true unless there is no
CP-violation in the mixing, i.e. $\dk =0$. We will describe Khalfin's result
in the next section. Before doing so let us state explicitly that in the
WW approximation we have $\pls = -\psln =0$ and that the $K_S$ and $K_L$ have
the simple time evolution
\begin{eqnarray} \label{2.14}
\pss  |_{{}_{WW}}= e^{-i m_{{}_{S}} t}e^{-{1 \over 2}
\Gamma_{{}_{S}} t} \nonumber \\
\pll |_{{}_{WW}}= e^{-i m_{{}_{L}} t}e^{-{1 \over 2}\Gamma_{{}_{L}} t}
\end{eqnarray}
as would have been expected for physical, unstable particle states (which do
not mix). As discussed above even in the WW approximation we have
\begin{equation} \label{2.15}
\mlst |_{{}_{WW}} \neq 0, \, \, \, \mslt |_{{}_{WW}} \neq 0
\end{equation}

It is also useful to derive two further relations which will be the
cornerstones of the discussion in
the next sections. The first one follows immediately from
eq.(\ref{2.12}) and reads
\begin{equation} \label{2.16}
\mslt + \mlst = \dk \left[ \mllt + \msst \right]
\end{equation}
This expression will lead in the next section to a relation between the
spectral density functions $\rho_{{}_{S}}$ and $\rho_{{}_{L}}$. This in turn
will yield a couple of consistency equation when the spectral functions are
approximated by a one-pole ansatz.
To obtain the second relation we have to essentially invert the formulae
(\ref{2.11}) and express the $\pkk$ etc matrix elements through the expansion
coeffients $\pss$ etc.
\begin{eqnarray} \label{2.17}
&\pkbk ={p \over q}\left\{{1 \over 2}
\left[\pss -\pll \right] + \pls \right\} \\
\label{2.18}
&\pbkk ={q \over p}\left\{{1 \over 2}
\left[\pss -\pll \right] - \pls \right\} \\
\label{2.19}
&\pkk =\pbkbk = {1 \over 2}\left[\pss + \pll \right]
\end{eqnarray}
Setting therein $\pls=0$ we get
\begin{equation} \label{2.20}
{\pkbk \over \pbkk} = {p^2 \over q^2}={\rm const}
\end{equation}
This last equation will, when rewritten in the spectral language, lead
to $\dk=0$. Hence the conclusion of Khalfin that $\psln \neq 0$.

\setcounter{equation}{0}
\section { Spectral Formulation}

What we called spectral formalism for unstable quantum mechanical systems is
based on two observations. The first one is simply the completeness of the
eigenvectors $|q\rangle$ of a hermitian quantum mechanical Hamiltonian. We can
then write an unstable state $|\lambda , t\rangle$ (which is never an
eigenstate of the Hamiltonian) as
\begin{equation} \label{3.1}
|\lambda , t\rangle =\sum_q |q,t\rangle \langle q|\lambda \rangle
\end{equation}
The second observation is the reasonable assumption that the unstable state has
only projections on continuum states in which it decays. Denoting from now on
the continous eigenvalue of a Hamiltonian by $m$ we can write the survival
amplitude $A(t)$ (or,
as in case of $K^0 \leftrightarrow \bar{K^0}$ oscillations, transition
amplitude) as
\begin{equation} \label{3.2}
A(t) = \int_{{\rm Spec}(H)}dm \, e^{-imt}\rho (m)
\end{equation}
where the integration extends over the whole spectrum of the Hamiltonian and
$\rho (m)$ is
\begin{equation} \label{3.3}
\rho (m) =|\langle m|\lambda \rangle |^2
\end{equation}
Of course the spectrum of any sensible Hamiltonian should be bounded from
below. The ground state (vacuum) can be then normalized to have zero energy
eigenvalue. The integration range in (\ref{3.2}) is in this case from
$0$ to $\infty$. Despite this cut-off in the integral (\ref{3.2}) imposed on
us by physical requirements we stress that $A(t)$ and $\rho (m)$ are still
Fourier-transforms of each other. This is guaranteed by the Dirichlet-Jordan
(see e.g. \cite{fourier})
conditions for Fourier integrals which under certain conditions (which we
assume here to be fullfiled) allow us to introduce a finite number of
discontinuities in the Fourier integrals. At the discontinuous points the
result of the Fourier transform will be $1/2[f(x+0)+f(x-0)]$ and simply $f(x)$
otherwise. \footnote{The other above mentioned conditions are (a) piecewise
continuity (except at isolated points), (b) bounded total variation and (c)
$\int_{-\infty}^{\infty}dt|A(t)|<\infty$. It is then sufficient to define
$\rho(m)\neq 0$ for $m \ge 0$ and $\rho(m)=0$ for $ m <0$. The absolute
integrability is obvious.}
With the following Breit-Wigner ansatz (see \cite{bohm})
\begin{equation} \label{3.4}
\rho_{{}_{BW}}(m)
= {\Gamma \over 2 \pi}\,{1 \over (m-m_0)^2 + {\Gamma^2 \over
4}}
\end{equation}
we obtain then for the survival amplitude
\begin{equation} \label{3.5}
A_{{}_{BW}}(t)
= \int^{\infty}_{-\infty} dm \, e^{-imt}\rho_{{}_{BW}}(m) =
e^{-im_0 t}\,e^{-{1 \over 2}\Gamma t},\, \, \, \, t \ge 0
\end{equation}
which gives for the survival probability the well known exponential decay law,
$P_{{}_{BW}}(t)=|A(t)|^2=\exp (-\Gamma t)$. Despite of what has been said about
the integration range above we have integrated in (\ref{3.5}) over
$(-\infty,\infty)$ for reasons which will be evident in section 5.
There it will
become apparent that taking the integral from $-\infty$ to $\infty$ is in some
sense equivalent to neglecting
terms of order $\Gamma /M$ (where $M$ is the mass).
The existence of a ground state in ${\rm Spec}(H)$ indroduces
non-exponential corrections (and non-oscillatory terms in $\pkk$ etc.)
which, however, using the simple ansatz (\ref{3.4}) cannot be trusted
\cite{bohm}. We will discuss this ansatz further in section 4.

We can now apply the above formalism to the case of $K_S$ and $K_L$ by
introducing a hermitian Hamiltonian with, as before, continuous spectrum of the
decay products which we label by indices $\alpha, \beta$ etc.
\begin{equation} \label{3.6}
H |\phi_{\alpha}\rangle = m |\phi_{\alpha} \rangle , \, \, \,
\langle \phi_{\beta}(m')|\phi_{\alpha}(m)\rangle =\delta_{\alpha \beta}
\delta (m' - m)
\end{equation}
The unstable states $K_S$ and $K_L$ are then written in accordance with
(\ref{3.1}) as superpositions of the eigenkets.
\begin{eqnarray} \label{3.7}
\ks = \int_0^{\infty} dm \, \sum_{\alpha}\rho_{{}_{S,\alpha}}(m)|\phi_{\alpha}
\rangle \nonumber \\
\kl = \int_0^{\infty} dm \, \sum_{\beta}\rho_{{}_{L,\beta}}(m)|\phi_{\beta}
\rangle
\end{eqnarray}
Note that this can be done for any unstable state. Therefore, strictly
speaking, equations (\ref{3.7}) are as such not the definitions of $\ks$ and
$\kl$. The latter are still defined as linear superposition of $|K^0 \rangle$
and $|\bar{K^0}\rangle$ in eq.(\ref{2.1}).

In what follows we convert the general formulae of section 2 into the langauge
of spectral functions $\rho (m)$. To do so we first write down the matrix
elements from eq.(\ref{2.12}). Using (\ref{3.6}) and (\ref{3.7}) they are
given by
\begin{eqnarray} \label{3.8}
& \msst = \int_0^{\infty} dm\, \sum_{\alpha}
|\rho_{{}_{S,\alpha}}(m)|^2 e^{-imt}  \nonumber \\
& \mllt = \int_0^{\infty} dm\, \sum_{\beta}
|\rho_{{}_{L,\beta}}(m)|^2 e^{-imt}   \nonumber \\
& \mslt = \int_0^{\infty} dm\, \sum_{\gamma}
\rho_{{}_{S,\gamma}}^* (m)\rho_{{}_{L,\gamma}}(m) e^{-imt} \nonumber \\
& \mlst = \int_0^{\infty} dm\, \sum_{\sigma}
\rho_{{}_{L,\sigma}}^*(m)\rho_{{}_{S,\sigma}}(m) e^{-imt}
\end{eqnarray}
Eq.(\ref{2.16}) can be then recast in the following form
\begin{eqnarray} \label{3.9}
\int_0^{\infty}dm \, \sum_{\alpha}
\left[\rho_{{}_{L,\alpha}}^*(m)\rho_{{}_{S,\alpha}}(m) +
\rho_{{}_{S,\alpha}}^*(m)\rho_{{}_{L,\alpha}}(m) \right]e^{-imt} \nonumber \\
= \dk \int_0^{\infty}dm\, \sum_{\beta}
\left[|\rho_{{}_{L,\beta}}(m)|^2 + |\rho_{{}_{L,\beta}}(m)|^2
\right]e^{-imt}
\end{eqnarray}
Taking the inverse Fourier transform of (\ref{3.9}) we arrive at
\begin{equation} \label{3.10}
\sum_{\alpha}
\left[\rho_{{}_{L,\alpha}}^*(m)\rho_{{}_{S,\alpha}}(m) +
\rho_{{}_{S,\alpha}}^*(m)\rho_{{}_{L,\alpha}}(m) \right] \nonumber
= \dk \sum_{\beta}
\left[|\rho_{{}_{L,\beta}}(m)|^2 +
|\rho_{{}_{L,\beta}}(m)|^2 \right]
\end{equation}
which is valid for $m \in (0, \infty )$. This equation is one of Khalfin's
main results \cite{kha1}
and will play an important role in the subsequent discussion. It
tells us that the spectral functions $\rho_{{}_{S,\alpha}}$ and
$\rho_{{}_{L,\alpha}}$ are inter-related with each other and any reasonable
ansatz which approximates these functions should be such that eq.(\ref{3.10})
is true at least to certain accuracy. Indeed an ansatz for
$\rho_{{}_{S,\alpha}}$ and
$\rho_{{}_{L,\alpha}}$ similar to (\ref{3.4}) does not fulfill this
requirements in full generality and in section 4 we address this question in
more detail. Note also that since eq.(\ref{3.10}) is an equation in the
variable $m$ we might expect that given a certain ansatz for the spectral
functions we get more than one consistency equations from it.

To obtain the second main result of Khalfin \cite{kha2}, \cite{kha3}
it is necessary to derive
corresponding spectral expression for $\pkk$ etc. From (\ref{2.12}),
(\ref{2.17})-(\ref{2.19}), (\ref{2.16}) (alternatively (\ref{3.10})) and
(\ref{3.8})we see that
\begin{eqnarray} \label{3.11}
\pkk &=& \pbkbk  =\int_0^{\infty}dm \, \rho_{{}_{K^0 K^0}}(m) e^{-imt}
\nonumber \\
\quad \quad &=& {1 \over
  2}\int_0^{\infty}\sum_{\alpha}\biggl\{|\rho_{{}_{S,\alpha}}(m)|^2 +
|\rho_{{}_{L,\alpha}}(m)|^2 \biggr\} e^{-imt} \\
\label{3.12}
\pkbk &=& \int_0^{\infty} dm \,\rho_{{}_{K^0 \bar{K^0}}}(m) e^{-imt} =
{1 \over 4p^* q}\int_0^{\infty}dm \, \sum_{\beta}\biggl\{|\rho_{{}_{S,\beta}
}(m)|^2
- |\rho_{{}_{L,\beta}}(m)|^2 \nonumber \\
& - & \rho_{{}_{S,\beta}}^*(m) \rho_{{}_{L,\beta}}(m) +
\rho_{{}_{L,\beta}}^*(m)\rho_{{}_{S,\beta}}
(m)\biggr\}e^{-imt} \\
\label{3.13}
\pbkk & =& \int_0^{\infty} dm \,\rho_{{}_{\bar{K^0} K^0}}(m) e^{-imt} =
{1 \over 4p q^*}\int_0^{\infty}dm \, \sum_{\sigma}\biggl\{|
\rho_{{}_{S,\sigma}}(m)|^2
- |\rho_{{}_{L,\sigma}}(m)|^2 \nonumber \\
& + & \rho^*_{{}_{S,\sigma}}(m) \rho_{{}_{L,\sigma}}(m) -
\rho^*_{{}_{L,\sigma}}(m)
\rho_{{}_{S,\sigma}}
(m)\biggr\}e^{-imt}
\end{eqnarray}
Here $\rho_{{}_{K^0 K^0}}(m)$ etc. are simply defined by the right
hand sides of
the corresponding equations. As done at the end of the forgoing section if
we now
set $\pls =-\psln =0$ we obtain the spectral version of (\ref{2.20})
\begin{equation} \label{3.14}
\int_0^{\infty}dm \, \rho_{{}_{K^0\bar{K^0}}}(m)e^{-imt} =
{p^2 \over q^2}\int_0^{\infty}dm \, \rho_{{}_{\bar{K^0}K^0}}(m)e^{-imt}
\end{equation}
By observing from (\ref{3.12}) and (\ref{3.13}) that $\rho_{{}_{K^0\bar{K^0}}}
=\rho^*_{{}_{\bar{K^0}K^0}}$ and taking again the inverse
Fourier transform in (\ref{3.14}) we get
\begin{equation} \label{3.15}
{p^2 \over q^2} = {\rho_{{}_{K^0\bar{K^0}}} \over \rho^*_{{}_{K^0 \bar{K^0}}}}
\end{equation}
This, however, immediately leads to
\begin{equation} \label{3.16}
\dk = |p|^2 - |q|^2=0
\end{equation}
Hence Khalfin's second result states that putting $\pls =-\psln$ to zero
invariably implies that on consistency grounds
there can be no CP-violation in the mixing provided the $K_S$ and $K_L$ states
are defined as in eqs.(\ref{2.1}). In other words since we know that
CP-violation exists in the mixing of $K^0-\bar{K^0}$ we have to allow for {\it
vacuum} regeneration of $K_S$ and $K_L$. Note that this conclusion does not
depend on a particular choice of $\rho_{{}_{S,\alpha}}$ and
$\rho_{{}_{L,\alpha}}$.
This is quite an astounding and
unexpected result which, using a completely
different approach, has also been recently confirmed \cite{suder3}.
It is not easy to give an
interpretation of this result. Either we accept (\ref{2.1}) and the fact that
the non-orthogonality of $K_S$ and $K_L$
makes this system different from
any other (recall our discussion of this peculiarity in the introduction) known
system (except for similar system with the same properties like
$B^0-\bar{B^0}$ or $D^0-\bar{D^0}$) or we can suspect that (\ref{2.1}) is not
the complete relation \cite{suder2}. The confirmation of the above
result by Chiu and
Sudershan \cite{suder3}
shows that this is result is indeed reliable. We emphasize this because of its
rather `exotic' implications.

It is also worthwhile noting that the above result has been derived within the
context of standard Quantum Mechanics and that CPT-symmetry has been
implemented. Suggested tests of CPT and Quantum Mechanics based
on terms which are in general forbidden by CPT or QM are then not affected by
this result provided the chosen observables assume zero values in the limit of
CPT conservation or in the context of QM. Any other tests which rely on
standard WW expressions might, however, be affected. This is true regardless
of the size of this new effect and more importantly
this effect has nothing to
do with deviations of the exponential decay law for very small and very large
time. The latter will become manifest in the formulae for time evolution in
section 5.

It is nevertheless mandatory to try to estimate the size of this effect. A
first step in this direction will be to make an ansatz for the spectral
functions $\rho_{{}_{S,\alpha}}$ and $\rho_{{}_{L,\alpha}}$
and to check the consistency of
this ansatz. Therefore we collect below all available expressions which can
shed some light on the spectral functions. From (\ref{2.1})-(\ref{2.3}) we
get
\begin{eqnarray} \displaystyle \label{3.17}
& \int_0^{\infty}dm \, \sum_{\alpha} |\rho_{{}_{S,\alpha}}(m)|^2 =
\int_0^{\infty}dm \, \sum_{\beta} |\rho_{{}_{L,\beta}}(m)|^2=1  \\
\label{3.18}
& \int_0^{\infty}dm \, \sum_{\sigma}\Im m \left(\rho^*_{{}_{S,\sigma}}(m)
\rho_{{}_{L,\sigma}}(m)\right)=0 \\
\label{3.19}
& \int_0^{\infty}dm \, \sum_{\gamma} \Re e
\left(\rho^*_{{}_{S,\gamma}}(m)
\rho_{{}_{L,\gamma}}(m)\right)=\dk
\end{eqnarray}
Eqs.(\ref{3.18}) and (\ref{3.19}) follow from (\ref{2.3}) and the fact that
$\dk$ is real. Together with (\ref{3.10}) these equations is all the
information on spectral functions $\rho_{{}_{S,\alpha}}$ and
$\rho_{{}_{L,\alpha}}$
which is given to us. Any ansatz for the spectral functions has to respect
these relations, up to a reasonable accuracy. We already mention that
Khalfin in his estimate (see also \cite{suder3}
where Khalfin's results and estimate are discussed)
used essentially only eq.(\ref{3.17}). We also point out that once
eq.(\ref{3.10}) and (\ref{3.17}) are assumed to hold eq.(\ref{3.19}) follows.
\setcounter{equation}{0}
\section{ One-Pole Approximation and its Consistency}

We have seen that the Breit-Wigner ansatz led to the well known exponential
decay law (up to corrections induced by the existence of the ground state). It
is therefore reasonable to assume a similar form for the $\rho_{{}_{S,\alpha}}$
and $\rho_{{}_{L,\alpha}}$. More specifically we write
\begin{eqnarray} \label{4.1}
 \rho_{{}_{S,\beta}}(m)=\sqrt{{\Gamma_{{}_{S}} \over 2 \pi}}\,
{A_{{}_{S,\beta}}(K_S
\to \beta) \over m - \ms +i{\Gamma_{{}_{S}} \over 2}} \nonumber \\
 \rho_{{}_{L,\beta}}(m)=\sqrt{{\Gamma_{{}_{L}} \over 2 \pi}}\,
{A{{}_{L,\beta}}(K_L
\to \beta) \over m - \ml +i{\Gamma_{{}_{L}} \over 2}}
\end{eqnarray}
where $A_{{}_{S,\alpha}}$ and $A_{{}_{L,\alpha}}$ are decay amplitudes. It is
convenient to make the following definitions
\begin{eqnarray} \label{4.2}
& \gsh \equiv {\Gamma_{{}_{S}} \over 2}, \, \, \,
\glh \equiv {\Gamma_{{}_{L}} \over 2}
, \, \, \, \dm \equiv \ms - \ml \\
\label{4.3}
& S \equiv \sum_{\alpha} |A_{{}_{S,\alpha}}|^2, \, \, \,
L \equiv \sum_{\alpha} |A_{{}_{L,\alpha}}|^2 \\
\label{4.4}
& R \equiv \sum_{\sigma} \Re e
\left(A^*_{{}_{S,\sigma}}A_{{}_{L,\sigma}}\right)
, \, \, \,
I \equiv \sum_{\sigma} \Im m \left(A^*_{{}_{S,\sigma}}A_{{}_{L,\sigma}}\right)
\end{eqnarray}
The quantities (\ref{4.3}) and (\ref{4.4}) are the only apriori unknown
variables which, with the spectral functions given by (\ref{4.1}), will enter
e.g. equations like (\ref{3.11})-(\ref{3.13}). As already mentioned at the end
of the last section we have to insert (\ref{4.1}) into the expressions
(\ref{3.10}) and (\ref{3.17})-(\ref{3.19}) to examine the consistency of the
one-pole approximation (\ref{4.1}).

We start with eq.(\ref{3.17}) where the integral can be easily performed. The
result is
\begin{equation} \label{4.5}
S= 1+ {\gsh \over \pi \ms}+ {\cal O}((\gsh / \ms)^2), \, \, \,
L= 1+ {\glh \over \pi \ml}+ {\cal O}((\glh / \ml)^2)
\end{equation}
For reasons explained below we will keep, up to a certain point, terms of order
$\Gamma_{{}_{X}}/m_{{}_{X}}$. Since (\ref{3.10}) contains the variable $m$
pluggging the one-pole approximation (\ref{4.1}) in (\ref{3.10}) we get a
polynomial in the variable $m$ of
degree two which should be identically zero. Therefore coefficient of
each power in $m$ should be also zero.
Instead of one equation we have
three consistency equations.
\begin{eqnarray} \label{4.6}
& m^2 \left[2\sqrt{\gsh \glh} \cdot R - \dk  (\gsh \cdot S + \glh \cdot L)
\right]=0
\nonumber \\
& m \left[-2\sqrt{\gsh \glh}(\ml + \ms) \cdot R - 2\sqrt{\gsh \glh}(\gsh -\glh)
\cdot I + 2\dk (\gsh \ml \cdot S + \glh \ms \cdot L) \right] =0
\nonumber \\
& \delta_{SL} \equiv  \dk \left[\gsh \cdot S(\ml^2 + \glh^2 )
+\glh \cdot L (\ms^2 + \gsh^2 )\right]
- 2\sqrt{\gsh \glh} (\gsh \glh + \ms \ml )\cdot R \nonumber \\
& +2 \sqrt{\gsh \glh} (\glh \ms
- \gsh \ml )\cdot I =0
\end{eqnarray}
{}From the first two we easily get
\begin{eqnarray} \label{4.7}
& R & = {\dk \over 2 \sqrt{\gsh \glh}} \left(\gsh \cdot S + \glh \cdot L
\right) \\
\label{4.8}
& I & = {\dk \over 2\sqrt{\gsh \glh}}\, {\dm \over \glh -\gsh}\left
(\gsh \cdot S -\glh \cdot L \right)
\end{eqnarray}
whereas the last condition in (\ref{4.6}) needs a more detailed treatment. The
reason why we did not neglect till now terms of order
$\Gamma_{{}_{X}}/m_{{}_{X}}$ is now apparent. Namely, in zeroth order of
$\Gamma_{{}_{X}}/m_{{}_{X}}$ we obtain
\begin{equation} \label{4.9}
\delta_{SL}|_{{}_{S=L=1}}=0
\end{equation}
Hence to estimate how badly $\delta_{SL}$ deviates from zero it is necessary to
include the next order of $\Gamma_{{}_{X}}/m_{{}_{X}}$. In this order
using (\ref{4.5})
$\delta_{SL}$ reads
\begin{eqnarray} \label{4.10}
& \delta_{SL}= {\dk \over \pi}\left({\glh \over \ml}\right){\gsh \over \gsh -
\glh}
\left[(\glh - \gsh )^2 +
\dm^2 \right]\left[(\gsh -\glh )- \gsh {\dm \over \ml}\right] \nonumber \\
& \sim  {\dk \over \pi} \dm^3 \left({\glh \over \ml }\right)
\end{eqnarray}
For the order of magnitude estimate in (\ref{4.10}) we have used
$\Gamma_{{}_{S}}/\dm \sim {\cal O}(1)$. Strictly speaking this amounts to
saying
that the ansatz (\ref{4.1}) is not consistent. Note, however, the following.
The smallest mass scale parameter which appears in calculations involving the
$K^0-\bar{K^0}$ system is $\dm$. $\delta_{SL}$ in (\ref{4.6}) has the canonical
dimension 3. What eq.(\ref{4.10}) then tells us is that as compared to the
third power of the smallest mass scale $\delta_{SL}$ is zero, up to corrections
of order $\Gamma_{{}_{X}}/m_{{}_{X}}$. Therefore to this accuracy everything is
consistent so far. Clearly, by assuming $\dk = 0$ we obtain $R=I=0$.

The reader will have noticed that in making estimates like in eq.(\ref{4.10})
we are relying on measured parameters of the $K^0-\bar{K^0}$ system. In order
not to lose track of the main point we will not examine simultaneuosly the
systems $B^0-\bar{B^0}$ and $D^0-\bar{D^0}$. There the smallest mass scale
parameter is not $\dm$ but the corresponding difference in the widths $\Delta
\Gamma$. The investigation of the consistency of (\ref{4.1}) will be then
slightly different in those systems. The general (hypothetical) case as well as
cases of physical interest other than the $K^0-\bar{K^0}$ system will be
treated elsewhere \cite{me}.

Using only Khalfin's eq.(\ref{3.10}) and the normalization condition
(\ref{3.17}) we have already pinned down the $S$, $L$, $R$ and $I$ in terms of
known quantities like widths, masses and $\dk$. The equations (\ref{3.10}) and
(\ref{3.17})-(\ref{3.19}) represent therefore
an over-determined system. In contrast to
situations discussed at the end of this section this is equivalent to a
consistency check.

On account of the validity of eq.(\ref{3.10}), proved for terms up to
$\Gamma_{{}_{X}}/m_{{}_{X}}$, eq.(\ref{3.19}) is bound to hold. We are
therefore
left with one more condition, namlely (\ref{3.18}). We will discuss the
calculation in connection with (\ref{3.18}) in some more detail since part of
the steps will enter also the formulae of time evolution in section 5. The
calculation will become more transparent by writing down explictly the product
$\sum_{\beta}\rho^*_{{}_{S,\beta}}(m)\rho_{{}_{L,\beta}}(m)$ with the spectral
functions given by (\ref{4.1}).
\begin{eqnarray} \label{4.11}
 \sum_{\beta}\rho^*_{{}_{S,\beta}}(m)\rho_{{}_{L,\beta}}(m)|_{{}_{BW}}
={\sqrt{\gsh \glh} \over \pi \left[(m-\ml)^2 + \gsh^2 \right]\left[(m-\ml)^2
+\glh^2 \right]} \nonumber \\
 \cdot \left\{(a_{{}_{R}}m^2+b_{{}_{R}}m + c_{{}_{R}}) +i(a_{{}_{I}}
m^2+ b_{{}_{I}}m + c_{{}_{I}})\right\}
\end{eqnarray}
with
\begin{eqnarray} \label{4.12}
& a_{{}_{I}}=I, \, \, \, \, b_{{}_{I}}=\left(\gsh -\glh \right) \cdot R -
\left(\ms + \ml \right)\cdot I \nonumber \\
&c_{{}_{I}}= \left(\glh \ms -\gsh \ml \right)\cdot R +
\left(\ml \ms + \gsh \glh \right) \cdot I
\end{eqnarray}
and similar expressions for $a_{{}_{R}}$, $b_{{}_{R}}$ and $c_{{}_{R}}$.
Next an ansatz for the  partial fraction decomposition
\begin{equation} \label{4.13}
{a_{{}_{I}}
m^2+ b_{{}_{I}}m + c_{{}_{I}} \over \left[(m-\ml)^2 + \gsh^2 \right]
\left[(m-\ml)^2 +\glh^2 \right]}= {C_I m + D_I \over (m-\ms )^2 +\gsh^2}
+ {E_I m + F_I \over (m-\ml )^2 +\glh^2}
\end{equation}
leads as usually to a linear system for coefficients $C_I$, $D_I$, $E_I$ and
$F_I$
\begin{eqnarray} \label{4.14}
&E_I =-C_I \nonumber \\
& C_I \dm + D'_I + F'_I = a_{{}_{I}} \nonumber \\
& C_I \left[\left(\ml^2 + \glh^2\right) -
\left(\ms^2 +\gsh^2 \right) \right] -2D'_I \ml -2F'_I \ms
=b_{{}_{I}} \nonumber \\
& D'_I \left(\ml^2 + \glh^2 \right) + F'_I \left(\ms^2 + \gsh^2 \right)
+ C_I \left[\ml\left(\ms^2 +\gsh^2 \right) -\ms \left(\ml^2 + \glh^2 \right)
\right]=c_{{}_{I}} \nonumber \\
\end{eqnarray}
where we have used the redefinitions
\begin{equation} \label{4.15}
D'_I \equiv D_I + C_I \ms , \, \, \, F'_I \equiv F_I - C_I \ml
\end{equation}
This system plays a double role in our discussion. It appears here as a middle
step in the consistency check and is a necessary ingredient in the calculation
of the time dependent transition amplitudes in the next section. Hence we feel
that it is of enough importance to give the explicit
solution of this system in appendix A. To perform the integral in (\ref{3.18})
we need also
\begin{eqnarray} \displaystyle \label{4.16}
&\Lambda (R,I)\equiv \int_0^{\infty}dm \, {a_{{}_{I}}
m^2+ b_{{}_{I}}m + c_{{}_{I}} \over \left[(m-\ml)^2 + \gsh^2 \right]
\left[(m-\ml)^2 +\glh^2 \right]}= \nonumber \\
& -C_I {\dm \over \ml} +
{D_I + C_I \ms \over \gsh} \left(\pi - {\gsh \over \ms}\right)
+ {F_I - C_I \ml \over \glh} \left(\pi - {\glh \over \ml}\right)
\nonumber \\
& +{\cal O}((\Gamma_{{}_{X}}/m_{{}_{X}})^2) +{\cal O}((\dm /\ml )^2)
\end{eqnarray}
such that the condition (\ref{3.18}) reduces to
\begin{equation} \label{4.17}
\Lambda (R,I)=0
\end{equation}
Taking the solutions for $C_I$, $D'_I$ and $F'_I$ in terms of $R$ and $I$
(see appendix A) and inserting them into (\ref{4.17}) a lengthy calculation
yields
\begin{eqnarray} \label{4.18}
&R \cdot  \dm \left[\dm^2 +\left(\gsh -\glh \right)^2 \right]\left[2\pi +
{\gsh + \glh \over \ml}\right] \nonumber \\
&+ I \cdot \left(\gsh + \glh \right)
\left[\dm^2 +\left(\gsh - \glh \right)^2 \right]
\left[2\pi - {\dm \over \ml}\, {\dm \over \gsh +\glh} \right]=0
\end{eqnarray}
In performing this calculation it is not advisable to make too strong
approximations right from the beginning. This is due to some
cancellations which
can occur. It is now trivial to compare eq.(\ref{4.18}) with (\ref{4.7}) and
(\ref{4.8}). In a simplified form eq.(\ref{4.18}) is
\begin{equation} \label{4.19)}
{I \over R}\simeq -{\dm \over \gsh +\glh}
+{\cal O}(\Gamma_{{}_{X}}/m_{{}_{X}}) +{\cal O}(\dm /\ml )
\end{equation}
which agrees with (\ref{4.7}) and (\ref{4.8}) when taking the approximation
$S=L \simeq 1$. \footnote{Yet a different way of displaying the consistency of
(\ref{3.18}) is descibed in section 5 (see there
eqs.(\ref{5.8})-(\ref{5.10})).}
The obvious conclusion here is that the one-pole ansatz
(\ref{4.1}) indeed passes the consistency check which has been imposed on us by
a set of equations in section 3. This check revealed that (\ref{4.1}) is valid
up to terms of order
${\cal O}(\Gamma_{{}_{X}}/m_{{}_{X}})$, ${\cal O}(\dm /\ml )$. We emphasize
that this is not a trivial check. To see this let us investigate the
situation where we put by hand $\dk =0$. In this case we would obtain an
homogeneous linear system whose only solution is $R=I=0$. No information on the
accuracy of (\ref{4.1}) would follow from this. On the other hand keeping
$\dk \neq 0$ but dropping Khalfin's eq.(\ref{3.10}) from the analysis we would
end up with four equations ((\ref{3.17})-(\ref{3.19}) for the four unkowns
$S$, $L$, $R$, and $I$. Again no conclusion on the accuracy could have been
reached. This displays once again the different nature of the $K^0-\bar{K^0}$
system and also the usefulness of (\ref{3.10}). As far as the {\it size } of
one possible correction term ($\sim \Gamma_{{}_{X}}/m_{{}_{X}}$) is concerned
the alert reader might object that this has been known all along as
corrections to the exponential decay law. This is only partly true. As we have
tried to argue above the presence of CP-violation alters the picture completely
as only in this case equations (\ref{3.10}) and (\ref{3.17})-(\ref{3.19})
are an overdetermined system. In this context we remark that:
1. a consistency check
has to be performed in any case as (\ref{4.1}) could have been
inconsistent for
totally different reasons and
2. it is probably safer not to rely on restrictions obtained
in the framework of a CP-conserving theory.
Corrections of the order ${\cal O}(\Gamma_{{}_{X}}/m_{{}_{X}})$ are
of course expected to the {\it exponential decay law}, but the result here is
more general as it explicitly states that corrections to {\it oscillatory
terms} in $\pkk$ etc. coming from exact (unkown) spectral functions
$\rho_{{}_{S,\alpha}}$ and $\rho_{{}_{L,\alpha}}$ will be of the same order.
Both these corrections are totally different in nature since corrections to
$\exp (-\Gamma t)$ are associated with the small/large time behaviour of the
amplitudes whereas corrections to oscillatory terms might also arise for
intermediate time scales. Indeed Khalfin's result on vacuum regeneration of
$K_S$ and $K_L$ discussed in section 3 induces corrections of the latter type
(see section 5). The nature of such corrections steming
from beyond (\ref{4.1}) cannot
be then apriori known and an analysis is required. That this analysis revealed
${\cal O}(\Gamma_{{}_{X}}/m_{{}_{X}})$ and ${\cal O}(\dm /\ml)$ as limits of
applicability of (\ref{4.1}) means also that we can trust terms of order ${\cal
O}(\Gamma_{{}_{L}}/\dm)$, should such terms indeed appear along the line of
further calculations. From now one we use
\begin{equation} \label{4.20}
S=L \simeq 1
\end{equation}
unless otherwise stated.

We close this section by observing that the sum of (\ref{4.7}) and (\ref{4.8})
with $S=L \simeq 1$ is nothing else but the well known Bell-Steinberger
unitarity relation \cite{bell2}, namely

\begin{equation} \label{4.21}
\dk \left(\gsh +\glh -i \dm \right) =2\sqrt{\gsh \glh} \sum_{\beta}
A^*_{{}_{S,\beta}}A_{{}_{L,\beta}}
\end{equation}
The reason it appears here in a slightly different form (compare e.g. with
\cite{maiani}) is the different normalization of the amplitudes. Recently
corrections
to (\ref{4.21}) of the order ${\cal O}(\dm /\ml)$  have been calculated (see
the second reference in \cite{suder2}). As shown above such corrections are
indeed expected. Finally we note that for the analysis in this section it is
immaterial whether ot not $\pls$ is zero.

\setcounter{equation}{0}
\section{ Time Development}

Having convinced ourselves that the one-pole approximation (\ref{4.1}) is
consistent up to terms of order
${\cal O}(\Gamma_{{}_{X}}/m_{{}_{X}})$ and ${\cal O}(\dm /\ml)$ we can proceed
to calculate the matrix elements (\ref{3.11})-(\ref{3.13}). With equations
(\ref{4.7}), (\ref{4.8}) and (\ref{4.20}) we have all necessary information to
do so. We mentioned in section 3 that the ground state in ${\rm Spec}(H)$
induces corrections to the exponential decay law (\ref{3.5}). Since this also
implies the integration domain $(0, \infty)$ in  (\ref{3.11})-(\ref{3.13})
we should handle such terms with care and make sure that all `new' terms
induced by the lower integration limit are indeed of strictly
non-oscillatory type in
(\ref{3.11})-(\ref{3.13}). This is also important as we want to find out if
Khalfin's effect is correlated with small/large time scales.
The relevant integrals have been calculated
analytically in appendix B. We can infer from the expressions in appendix B
that such terms contain the
exponential integral function $Ei$ \cite{grad}.
We can safely neglect the terms with $Ei$
as it should be clear that the simple ansatz (\ref{4.1}) cannot account for the
correctness of such non-oscillatory terms.

Let us now have a closer look at (\ref{3.11}). In the one-pole approximation
(\ref{4.1}) $\pkk$ can be conveniently written as (see also (\ref{B9}) in
appendix B)
\begin{eqnarray} \label{5.1}
\pkk = \pbkbk& =&{1 \over 2 \pi}\biggl\{
e^{-im_{{}_{S}} t}\left(-\int_0^{-m_{{}_{S}} /\gamma_{{}_{S}}} dy
\, {e^{-i\gamma_{{}_{S}} ty} \over y^2 + 1} +\int_0^{\infty}dy \,
{e^{-i \gamma_{{}_{S}} ty} \over
y^2 +1}\right) \nonumber \\
&+ &\left[S \to L \right]\biggr\}
\end{eqnarray}
We see that we have to calculate integrals of the following type
\begin{eqnarray} \label{5.2}
K^{(n)}(a) &\equiv & \int_0^{\infty} dx \, {x^n \over x^2 +1}e^{-iax}
\nonumber \\
J^{(n)}(a,\eta) & \equiv & \int_0^{\eta}dx \, {x^n \over x^2 + 1}e^{-iax}
\end{eqnarray}
Collecting only oscillatory terms from the integrals in appendix B  we obtain
the same expression as in WW-approximation (this of course is not a surprise
recalling that our concern here is the last equation in (\ref{2.11}) where
only $\pkbk$ and $\pbkk$ play a role)
\begin{equation} \label{5.3}
\pkk = \pbkbk ={1 \over 2} \left\{e^{-im_{{}_{S}} t}e^{-\gamma_{{}_{S}} t}
+ e^{-im_{{}_{L}} t}e^{-\gamma_{{}_{L}} t}\right\} + N_{{}_{K^0 K^0}}(t)
\end{equation}
where $N_{{}_{K^0 K^0}}(t)$ denotes all non-oscilllatory terms present in the
integral. $N_{{}_{K^0 K^0}}(t)$ can, in principle, be extracted from equations
(\ref{B1})-(\ref{B5}) but as we said before we cannot trust such terms to be
the correct non-oscillatory corrections.

One more comment is order. Putting $\gsh /\ms$ to zero the sum of the two
integrals in (\ref{5.1})
can be compactly written as
\begin{equation} \label{5.4}
\int_{-\infty}^{\infty}dy \, {e^{-i\mu y} \over a^2 + y^2}={\pi \over a}
e^{-\mu a}
\end{equation}
which of course means that
\begin{equation} \label{5.5}
N_{{}_{K^0 K^0}}(t) \to 0 \, \, \, {\rm as}\, \, \,
{\Gamma_{{}_{S/L}} \over m_{{}_{S/L}}}\to 0
\end{equation}
in agreement with what we said at the beginning of section 3 (see discussion
below eq.(\ref{3.5})).

Similarly the integration in (\ref{3.12}) and (\ref{3.13}) can be done
analytically (see (\ref{B10}) in appendix B) and the result reads
\begin{eqnarray} \label{5.6}
\pkbk &=&{1 \over 4p^* q}\left\{e^{-im_{{}_{S}} t}e^{-\gamma_{{}_{S}} t}
[1+\kappa_{{}_{S}}]- e^{-im_{{}_{L}} t}e^{-\gamma_{{}_{L}} t}
[1+\kappa_{{}_{L}}]\right\} +N_{{}_{K^0 \bar{K^0}}}(t)
\nonumber \\
\pbkk &=&{1 \over 4p q^*}\left\{e^{-im_{{}_{S}} t}e^{-\gamma_{{}_{S}} t}
[1-\kappa_{{}_{S}}]- e^{-im_{{}_{L}} t}e^{-\gamma_{{}_{L}} t}
[1-\kappa_{{}_{L}}]\right\} +N_{{}_{\bar{K^0} K^0}}(t)
\end{eqnarray}
where $N_{{}_{K^0 \bar{K^0}}}(t)$ and $N_{{}_{\bar{K^0} K^0}}(t)$ are again
non-oscillatory terms containing the exponential integral function $Ei$ and
$\kappa_{{}_{S/L}}$ are given by
\begin{eqnarray} \label{5.7}
\kappa_{{}_{S}}&=&-2i{\sqrt{\gsh \glh} \over \gsh}\left[D'_I -i\gsh C_I \right]
\nonumber \\
\kappa_{{}_{L}}&=&+2i{\sqrt{\gsh \glh} \over \glh}\left[F'_I
+i\glh C_I \right]
\end{eqnarray}
The parameter $C_I$, $D'_I$ and $F'_I$ are defined as solutions of the linear
system (\ref{4.14}). Equation (\ref{5.6}) together with (\ref{2.11}) shows
that Khalfin's effect depends crucially
on the size of the quantities $\kappa_{{}_{S/L}}$. We could, in principle,
calculate these quantities taking the solutions $C_I$, $D'_I$ and $F'_I$ from
appendix A. There is, however, a more elegant way by going back to
(\ref{4.14}).
This linear system fixes $C_I$, $D'_I$ and $F'_I$ in terms of $R$ and $I$
(eqs.(\ref{4.7})-(\ref{4.8})) the latter being kept at this stage in section 4
arbitrary i.e. in any order of $\Gamma_{{}_{X}} /m_{{}_{X}}$.
But we know now that we are allowed to keep only the
zeroth order of $\Gamma_{{}_{X}} /m_{{}_{X}}$. Then $R$, $I$ taken
together with  (\ref{4.20}) and a redefinition of the form
\begin{equation} \label{5.8}
\left(
\begin{array}{c}
 \tilde{C}_I  \\
 \tilde{D}_I \\
 \tilde{F}_I
\end{array}
\right)=
2{\sqrt{\gsh \glh} \over \dk}
\left(
\begin{array}{c}
 C_I \\
 D'_I \\
 F'_I
\end{array} \right)
+ \left(
\begin{array}{c}
 1 \\
 0 \\
 0
\end{array}\right), \, \, \,
\left(
\begin{array}{c}
 \tilde{a}_{{}_{I}}  \\
 \tilde{b}_{{}_{I}}  \\
 \tilde{c}_{{}_{I}}
\end{array} \right)
=2 {\sqrt{\gsh \glh} \over \dk}
\left(
\begin{array}{c}
 a_{{}_{I}}  \\
 b_{{}_{I}}  \\
 c_{{}_{I}}
\end{array} \right)
\end{equation}
convert (\ref{4.14}) into a homogeneous linear
system in the limit $\Gamma_{{}_{X}}/m_{{}_{X}} \to 0$
\begin{equation} \label{5.9}
\left( \begin{array}{ccc}
-\tilde{a}_{{}_{I}} & 1 & 1 \\
-\tilde{b}_{{}_{I}} & -2\ml & -2\ms \\
-\tilde{c}_{{}_{I}} & \ml^2 + \glh^2 & \ms^2 +\gsh^2
\end{array} \right)
\left( \begin{array}{c}
\tilde{C}_I \\
\tilde{D}_I \\
\tilde{F}_I
\end{array} \right)=0
\end{equation}
Since the determinant \footnote{Demanding the determinant to be zero gives
$\tilde{b}_{{}_{I}}-\tilde{a}_{{}_{I}}\tilde{c}_{{}_{I}}=0$ and in the suitable
accuracy
$\gsh^2 -\glh^2 \simeq 0$ or $(\gsh^2 -\glh^2)/\dm^2 \simeq -2(\ms +\ml )/
\dm$. Clearly both these relations are only hypothetical and not valid in the
$K^0-\bar{K^0}$ system.}
of the cofficient matrix in (\ref{5.9}) is non-zero we
get only a trivial solution
\begin{equation} \label{5.10}
\tilde{C}_I=\tilde{D}_I=\tilde{F}_I=0
\end{equation}
This immediately implies that
\begin{equation} \label{5.11}
\kappa_{{}_{S}}=\kappa_{{}_{L}}=\dk +{\cal O}(\Gamma_{{}_{X}}/m_{{}_{X}})
+{\cal O}(\dm /\ml )
\end{equation}
Equipped with this simple result eq.(\ref{5.6}) take the familar form
\begin{eqnarray} \label{5.12}
\pkbk = {p \over 2q}\left\{e^{-im_{{}_{S}} t}
e^{-\gamma_{{}_{S}} t}- e^{-im_{{}_{L}} t}e^{-\gamma_{{}_{L}} t}
\right\} + {\rm non-osc.\, terms} \nonumber \\
\pbkk = {q \over 2p}\left\{e^{-im_{{}_{S}} t}
e^{-\gamma_{{}_{S}} t}- e^{-im_{{}_{L}} t}e^{-\gamma_{{}_{L}} t}
\right\} + {\rm non-osc.\, terms}
\end{eqnarray}
Up to non-oscillatory terms these equations are equivalent to the
WW-expressions.
What we have shown is that indeed corrections
to oscillatory terms due to Khalfin's general result
will appear in (\ref{5.12}), but they are necessarily of the order
${\cal O}(\Gamma_{{}_{X}}/m_{{}_{X}})$, ${\cal O}(\dm /\ml )$. This follows
from the fact that the one-pole approximation is trustable only up to such
terms. In the calculation with the one-pole ansatz
any term whose order of magnitude is much bigger than
${\cal O}(\Gamma_{{}_{X}}/m_{{}_{X}})$, ${\cal O}(\dm /\ml )$, like
$\Gamma_{{}_{L}}/\dm$, would be then still acceptable. But such a term does not
show up along the line of the calculation. It should also be appreciated that
such corrections have nothing to do with small/large time behaviour of the
transition amplitudes (i.e. they are not interrelated to the usual corrections
to the exponential decay law). This is evident from the way $\kappa_{{}_{S/L}}$
enters (\ref{5.6}).

Finally the  answer to the question we have put forward in the
form of equation
(\ref{2.13}) can also be given by a simple equation, namely
\begin{equation} \label{5.13}
\pls = -\psln =0
+{\cal O}(\Gamma_{{}_{X}}/m_{{}_{X}}) +{\cal O}(\dm /\ml )
\end{equation}
Had we not Khalfin's theorem discussed insection 3, it would be compeletely
legitimate to assume $\pls$ to be strictly zero.
Our result agrees with the conclusion of ref. \cite{suder3}
reached there in a different
way. We postpone any further discussion to the next section where we
will give a
summary. In the end we
compare our result with expressions obtained by Khalfin who arrives at a
equation similar to (\ref{5.6}) \cite{kha3}. To obtain his results we
have to make only the
following replacement in eq.(\ref{5.6})
\begin{equation} \label{5.14}
\kappa_{{}_{S}} \to {-2i \sqrt{\Gamma_{{}_{S}}\Gamma_{{}_{L}}}
\over \dm +i\left(\Gamma_{{}_{S}} + \Gamma_{{}_{L}}\right)},
\, \, \, \, \, \, \kappa_{{}_{L}} \to
\kappa^*_{{}_{S}}
\end{equation}
As explicitly shown in \cite{suder3}
the numenrical value of $\kappa_{{}_{S}}$ would
then be
\begin{equation} \label{5.15}
\kappa_{{}_{S}} \sim 0.06\, e^{i\pi /4}
\end{equation}
The effect would then indeed be of the order $\Gamma_{{}_{L}}/\dm$ as can be
seen from the equation
\begin{equation} \label{5.16}
 |\pkbk |^2 \simeq {1 \over 4}\left\{e^{-\Gamma_{{}_{S}}t}
+ e^{-\Gamma_{{}_{L}}t}
 -2e^{-(\gamma_{{}_{S}}+ \gamma_{{}_{L}})t}\left[\cos (\dm t) -
0.4 \times 10^{-3}\sin (\dm t)\right]\right\}
\end{equation}
We have, hoever, shown that this is an overestimate by several orders of
magnitude. The difference between Khalfin's approach and ours is essentially
our consistent treatment of the one-pole approximation in section 4.

\setcounter{equation}{0}
\section{Conclusions}

It is satisfactory to arrive after lengthy calculations at familar expressions
of the Weisskopf-Wigner approximation. More so as our starting point was
completely different from the WW-approach. This not only gives us more
confidence in the WW-approximation whose equations, as we know, are of utmost
importance for the $K^0-\bar{K^0}$ system, but has also the virtue that one is
able to derive the limitations of the WW-approximation for the oscillatory as
well as for the exponential terms. We have emphasized that corrections to the
oscillatory terms are different in nature from corrections arising from
small/large time behaviour of the amplitudes. It turned out, however, that both
such corrections must be of the order
${\cal O}(\Gamma_{{}_{X}}/m_{{}_{X}})$, ${\cal O}(\dm /\ml )$. This is apriori
not evident due to the specifics of the $K^0-\bar{K^0}$ system where
beside $\Gamma_{{}_{X}}/m_{{}_{X}}$ quantities
like $\Gamma_{{}_{L}}/\dm$ do appear. The reanalysis of the present paper was
also necessary in view of a claim of Khalfin that new effects in connection
with
the non-zero vacuum regeneration of $K_S$ and $K_L$ are of the order of
$\Gamma_{{}_{L}}/\dm$. Let us recapitulate the steps which have led to our
result. We have presented two of Khalfin's theorems. One was eq.(\ref{3.10})
which played a crucial role in our analysis. Actually without this equation no
conclusion on the validity of the one-pole approximation could have been
reached.
The other one was the surprising
result on the existence of $K_S$ and $K_L$ vacuum regeneration, an effect
usually associated with interactions of $K_S$ and $K_L$ in matter. Although
this result is quite `exotic' the author of the present paper could not find
a loop-hole in the arguments which led to this result. The vacuum regeneration
of $K_S$ and $K_L$ goes against what one would intuitively expect and what one
is normally used to. Note, however, the this `intuition' is based on quantum
mechanical systems where the unstable states have zero overlap. $\ks$ and $\kl$
have non-zero overlap, a singled-out property which is then responsible for
counter-intuitive effects. The proof of Khalfin's result relies on well
established formalims of Quantum Mechnics (eqs.(\ref{3.1})-(\ref{3.3})) and
seems therefore hard to dispute once we assume that
$\ks$ and $\kl$ are given as in
(\ref{2.1}).
To estimate the size of
such an effect we had to perform a consistency check of the one-pole
approximation (\ref{4.1}). The outcome of this check provided us with limits of
the applicability of (\ref{4.1}) and the determination of apriori unkown
variables (combinations of decay amplitudes). Indeed the difference between the
present paper and the result obtained by Khalfin can be traced back to exactly
this point. In a subsequent step we have derived the time evolution of the
system starting from the equations (\ref{3.11})-(\ref{3.13}). The formulae so
obtained agreed with expressions from the WW-formalism. This in turn implied
that the effect of vacuum regeneration of $K_S$ and $K_L$ is necessary small
and of the order of
${\cal O}(\Gamma_{{}_{X}}/m_{{}_{X}})$, ${\cal O}(\dm /\ml )$

Our estimate does not render the general result of Khalfin useless as in fact
the effect is non-zero. Furthermore we know from this result that on quite
general grounds
\begin{equation} \label{6.1}
{\pkbk \over \pbkk} \neq {\rm const}
\end{equation}
Any test therefore which as a starting assumption relies instead on
(\ref{2.20}) \cite{dass} should be then carefully reconsidered.

\vskip  2cm

{\elevenbf \noindent Acknowledgments}
\newline
The author thanks N. Paver, G. Pancheri, M. Finkemeier and
R.M. Godbole for useful discussions and comments on the subject treated in the
paper. The author also wishes to acknowledge financial support by the HCM
program under EEC contract number CHRX-CT 920026.
\vglue 0.4cm
\newpage
\setcounter{equation}{0}

\appendix
\section*{Appendix A}
\renewcommand{\theequation}{A.\arabic{equation}}

We list here the solutions of the linear system (\ref{4.14}). Since the
expressions are lengthy it is convenient to use the following notational
abbreviations
\begin{eqnarray} \label{A1}
& X_+ = \gsh^2 + \glh^2, \, \, \, X_- = \gsh^2 - \glh^2 \nonumber \\
& Z=\ml \gsh^2 -\ms \glh^2 \nonumber \\
& Y_I= \dm^4 +2 \dm^2 X_+ + X_-^2
\end{eqnarray}
The solutions in terms of $a_{{}_{I}}$, $b_{{}_{I}}$ and $c_{{}_{I}}$ defined
in eq.(\ref{4.12}) then read
\begin{eqnarray} \label{A2}
F'_I \cdot Y_I =& a_{{}_{I}}& \left[\dm^2 \ml^2 - (\ml^2 + \glh^2)X_- -\dm \ml
X_+ +(\ml +\ms)Z\right]+ \nonumber \\
&b_{{}_{I}}&\left[\dm^2 \ml +Z - \dm \glh \right]+ \nonumber \\
&c_{{}_{I}}&\left[\dm^2 +X_- \right]
\end{eqnarray}

\begin{eqnarray} \label{A3}
D'_I \cdot Y_I =& a_{{}_{I}}& \left[\dm^2 \ms^2 + (\ms^2 + \gsh^2)X_- +\dm \ms
X_+ -(\ml +\ms)Z\right]+ \nonumber \\
&b_{{}_{I}}&\left[\dm^2 \ms -Z + \dm \gsh \right]+ \nonumber \\
&c_{{}_{I}}&\left[\dm^2 -X_- \right]
\end{eqnarray}

\begin{eqnarray} \label{A4}
C_I \cdot Y_I =& a_{{}_{I}}& \left[\dm^3 -\dm(\ms^2+\ml^2-\gsh^2-\glh^2)
-(\ms +\ml)X_- \right] -\nonumber \\
&b_{{}_{I}}&\left[\dm (\ms +\ml)  + X_- \right]- \nonumber \\
&c_{{}_{I}}&2\dm
\end{eqnarray}

\setcounter{equation}{0}
\section*{Appendix B}
\renewcommand{\theequation}{B.\arabic{equation}}

This appendix contains the relevant integrals appearing in
(\ref{3.11})-(\ref{3.13}) with $\rho_{{}_{S,\alpha}}$ and
$\rho_{{}_{L,\alpha}}$
approximated by (\ref{4.1}). The integrals $K^{(n)}(a)$ and $J^{(n)}(a,\eta)$
are defined in (\ref{5.2}). We have \cite{grad}
\begin{equation} \label{B1}
K^{(0)}(a)=\int_0^{\infty}dx \, {1 \over x^2 +1}e^{-iax}={\pi \over 2}
e^{-a} - {i \over 2}\left[e^{-a}Ei(a) -e^a Ei(-a) \right]
\end{equation}
where $Ei$ are transzendental functions called exponential integral functions.
Any other integral $K^{(n)}$ with $n > 0$ can be obtained from (\ref{B1}) by
differentiating (\ref{B1}) with respect to the variable $a$. For instance
\begin{equation} \label{B2}
K^{(1)}(a)=\int_0^{\infty}dx \, {x \over x^2 +1}e^{-iax}=-i{\pi \over 2}
e^{-a} - {1 \over 2}\left[e^{-a}Ei(a) +e^a Ei(-a) \right]
\end{equation}
The integral $J^{(n)}$ are more complicated. To obtain $J^{(0)}$ we have used
the Fourier indentity
\begin{equation} \label{B3}
\int_0^{\eta}f(x)dx = {1 \over 2\pi}\int_{-\infty}^{\infty}dy \, {e^{i\eta y}
-1 \over iy}\int_{\infty}^{\infty}e^{-iy\xi}f(\xi)d\xi
\end{equation}
Here we quote only the result
\begin{eqnarray} \label{B4}
& J^{(0)}(a,\eta)=\int_0^{\eta}dx \, {1 \over x^2 +1}e^{-iax}=
-{1 \over 2i}\biggl\{-i{\rm sgn}(\eta)e^{-a} + e^{-a}Ei(a(1-i\eta))
\nonumber \\
&-e^a Ei(-a(1+i\eta)) - e^{-a} Ei(a) + e^{a}Ei(-a)\biggr\}
\end{eqnarray}
where ${\rm sgn}(\eta)$ is the sign of $\eta$. Again $J^{(n)}$, $n > 0$ can be
obtained from (\ref{B4}) by differentiation of (\ref{B4}) with respect to $a$
\begin{eqnarray} \label{B5}
& J^{(1)}(a,\eta)=\int_0^{\eta}dx \, {x \over x^2 +1}e^{-iax}=
-{1 \over 2}\biggl\{i{\rm sgn}(\eta)e^{-a} - e^{-a}Ei(a(1-i\eta))
\nonumber \\
&-e^a Ei(-a(1+i\eta)) + e^{a} Ei(-a) + e^{-a}Ei(a)\biggr\}
\end{eqnarray}
The reason why we have to distinguish between the signs of $\eta$ has to do
with the following property of the exponential integral function $Ei$
\cite{grad}
\begin{equation} \label{B6}
Ei(x \mp i0)=Ei(x) \pm i\pi, \, \, \, \, x>0
\end{equation}
One can check (\ref{B4}) by using the integral representation
\begin{equation} \label{B7}
Ei(\pm xy)=\pm e^{\pm xy}\int_0^{\infty}dt \, {e^{-xt} \over y \mp t},
\, \, \, \Re e(y) > 0, \, x>0
\end{equation}
and differentiating both sides of (\ref{B4}) with respect to $\eta$. We also
mention here the connection of $Ei(x)$ with the incomplete beta function
$\Gamma (\alpha , x)$ \cite{grad} through
\begin{equation} \label{B8}
\Gamma (0,x)=-Ei(-x)
\end{equation}
Finally the integrals (\ref{B1})-(\ref{B5}) enter (\ref{3.11})-(\ref{3.13})
through the expressions
\begin{equation} \label{B9}
\int_0^{\infty}dm \, \sum_{\alpha}|\rho_{{}_{S,\alpha}}(m)|^2 e^{-imt}={1 \over
\pi}e^{-im_{{}_{S}}t}\left[-J^{(0)}(\gsh t, -\ms / \gsh ) +K^{(0)}(\gsh
t)\right]
\end{equation}
and
\begin{eqnarray} \label{B10}
& \int_0^{\infty}dm \, \sum_{\beta} \Im
m\left(\rho_{{}_{S,\beta}}(m)\rho^*_{{}_{L,\beta}}(m)\right)
e^{-imt}= \nonumber \\
&-{\sqrt{\gsh \glh} \over \pi} \int_0^{\infty}dm \, {a_{{}_{I}}
m^2+ b_{{}_{I}}m + c_{{}_{I}} \over \left[(m-\ml)^2 + \gsh^2 \right]
\left[(m-\ml)^2 +\glh^2 \right]}e^{-imt}= \nonumber \\
& -{\sqrt{\gsh \glh} \over \pi}\biggl\{{e^{-im_{{}_{S}}} \over \gsh}\biggl[
D'_I\cdot
\left(-J^{(0)}(\gsh t, -\ms / \gsh )+K^{(0)}(\gsh t)\right) \nonumber \\
&+\gsh C_I\cdot \left(-J^{(1)}(\gsh t, -\ms / \gsh )+K^{(1)}(\gsh
t)\right)\biggr] \nonumber \\
&+{e^{-im_{{}_{L}}} \over \glh}\biggl[
F'_I\cdot
\left(-J^{(0)}(\glh t, -\ml / \glh )+K^{(0)}(\glh t)\right) \nonumber \\
&-\glh C_I\cdot
\left(-J^{(1)}(\glh t, -\ml / \glh )+K^{(1)}(\glh t)\right)\biggr]
\biggr\}
\end{eqnarray}
\newpage


\end{document}